\documentclass[dvips,12pt]{article}
\usepackage{graphicx}
\begin{document}
\hfill\vbox{\baselineskip14pt
            \hbox{\bf March 2006}}
\baselineskip20pt
\vskip 0.2cm
\begin{center}
{\Large\bf Clear Experimental Signature of Charge-Orbital density wave in
Nd$_{1-x}$Ca$_{1+x}$MnO$_{4}$:Heat Capacity and Magnetization study}
\end{center}
\begin{center}
\large Sher~Alam$^{1}$, A.T.M.N. Islam$^{2}$,~T.Nagai$^{1}$,~ M.~Xu$^{1}$,
Javed Ahmad$^{3}$, Y.~Matsui$^{1}$,~and ~I.~Tanaka$^{2}$
\end{center}
\begin{center}
$^{1}${\it AML, NIMS, Tsukuba 305-0044, Ibaraki, Japan}\\
$^{2}${\it Crystal Group, Yamanashi University, Yamanashi, Japan}\\
$^{3}${\it Dept. of Physics, Bahauddin Zakariya University, Multan, Pakistan}
\end{center}
\begin{center}
\large Abstract
\end{center}
\begin{center}
\begin{minipage}{16cm}
\baselineskip=18pt
\noindent
  Single Crystals of Nd$_{1-x}$Ca$_{1+x}$MnO$_{4}$ have been prepared by
the traveling floating-zone method, and possible evidence of a charge
-orbital density wave in this material presented earlier [PRB68,092405 (2003)]
using High Resolution Electron Microscopy [HRTEM] and Electron
Diffraction [ED]. In the current note we present direct evidence of
charge-orbital ordering in this material using heat capacity
measurements. Our heat capacity measurements indicate a clear transition
consistent with prior observation. We find two main transitions, one at
temperature $T_{_H}^{HC}=310-314$ K, and other in the vicinity of $T_{_A}^{HC}=143$ K.
In addition, we may also conclude that there is a strong electron-phonon
coupling in this material. In order to further study and confirm these
anomalies we have performed dc magnetization measurements. The dc magnetic
measurements confirm these two transitions. Again,we find two main transitions, one at
temperature $T_{_H}^{M}=318-323$ K, and other at around $T_{_A}^{M}=164$ K.
\end{minipage}
\newline
\vskip 0.2cm
PACS numbers: 78.20.-e, 78.30.-j, 74.76.Bz
\end{center}
\vfill
\baselineskip=20pt
\normalsize
\newpage
\setcounter{page}{2}
\section{Introduction}
The transition-metal oxides display a wide variety of interesting
properties. In particular of interest are the 3d transition metals
such as Fe, Cu, Ni, Co and Mn, having a single layered perovskite
structure, i.e. K$_2$NiF$_4$-type. These materials provide us useful
insights into the underlying Mott-Hubbard in addition to helping us
understand the superconducting oxides to which they are similar.
In a recent note we \cite{alam02} have shown the existence of
inhomogenous charge distribution in very good quality
single crystal of LaSrCuO$_{4}$ [LSCO] co-doped with $1\%$ Zn at
the copper site, using temperature dependent polarized X-ray absorption
near-edge structure [XANES] spectra. A related single-layered compound
to the LSCO is LaSrFeO$_{4}$ whose atomic and magnetic structure was
first elucidated by Souberoux et al.\cite{sou1980},
using X-ray diffraction, Mossbauer spectroscopy and neutron diffraction
in polycrystalline state. Recently Kawanaka \cite{kawa03} have
prepared  a single crystal of  LaSrFeO$_{4}$ by floating-zone
method and demonstrated a spin flop transition in the antiferromagnetic
ground state.

In particular it seems that 3d-electron systems exhibit ordering
and disordering of the fundamental degrees of freedom such as
charge, spin and orbital. In turn it is plausible that this
may be responsible for metal-insulator transition\cite{kawa03},
high-T$_c$ superconductivity \cite{alam02, alam02-0}, and colossal
magnetoresistance \cite{ima98}. A class of materials which show
charge-orbital ordering of $e_g$ electrons are the mixed valent
manganites, the perovskite-type: RE$_{1-x}$AE$_{x}$MnO$_{3}$,
single-layered RE$_{1-x}$AE$_{1+x}$MnO$_{4}$ and double-layered
 RE$_{2-2x}$AE$_{1+2x}$Mn$_{2}$O$_{7}$, where RE=trivalent lanthanides,
and AE=divalent alkaline-earth ions. The real-space ordering pattern,
possibly induced by strong electron-phonon interaction, that is
a cooperative Jahn-Teller effect, can be determined by crystallographic
superstructure \cite{nag03, nag02}. What physical insight can be gained
from these crystallographic superstructure? Here are some examples,
for the case of the over-doped single-layered manganites, the suggested
models\cite{nag02, lar01} indicate that the Mn valence is not an integer
but suffers a modulation from site to site. In addition two other
theoretical proposals of some interest for the over-doped single-layered
manganites, are the Wigner-crystal\cite{che97} and bi-stripe
models\cite{mor98}.
The Wigner-crystal model \cite{che97} asserts the stacking of
Mn$^{3+}$ and Mn$^{4+}$ stripes, such that  Mn$^{3+}$ stripes are
regularly spaced, in contrast the bi-stripe model \cite{mor98} claims
the pairing of the Mn$^{3+}$ stripes.

Thus main the purpose of this note is to concentrate on the
experimental results of our heat capacity measurements for the
layered manganites Nd$_{1-x}$Ca$_{1+x}$MnO$_{4}$. Previously two
of us reported \cite{nag03}, using HRTEM and ED presence of
super-reflections and transverse and sinusoidal modulations
in the charge-orbital ordered phases. The previous measurements \cite{nag03}
were carried out using polycrystalline samples of
Nd$_{1-x}$Ca$_{1+x}$MnO$_{4}$. Here we report for the first time the
successful growth of single crystals of Nd$_{1-x}$Ca$_{1+x}$MnO$_{4}$
for x=0.67 and x=0.70. The Heat Capacity [HC] was measured using
Quantum Design Physical Property Measurement System [PPMS].

The dc magnetization, was measured using Quantum Design SQUID
magnetometer [MPMS].

This paper is organized as follows, in the
next section we outline experimental details. This is followed
by section three where the results and discussion of our study
are given. The final section contains the conclusions.


\section{Experimental}

It should be noted that we have for the first time succeeded in growing
single crystals of Nd$_{1-x}$Ca$_{1+x}$MnO$_{4}$ for x=0.67 and x=0.70.
It is non-trivial to do so, in particular for the case of high calcium
content, i.e. $x \geq 0.72$ it was not possible to obtain single crystals.

The crystal growth was carried out using the floating zone technique in a
four-mirror type infrared image furnace. Feed rod was prepared by solid state
reactions: high purity powders of Nd$_2$O$_3$, CaCO$_3$ and MnO$_2$ were
mixed in stoichiometric composition and were calcined twice at
1000 $^{o}$C for
12 hours with intermediate grinding. In fact about 3\% extra MnO$_2$ was
mixed to compensate for the vaporization. The powder was then packed in
a rubber tube and cold pressed under 300 MPa of pressure to a dense rod
of ~6 mm in diameter and 50-70 mm in length. The feed rod was then sintered
at 1500 $^{o}$C for 12 hours. Crystal growth was performed under flowing
oxygen atmosphere at a speed of 5-10 mm/h.

\begin{figure}
\includegraphics[width=6in,height=6in]{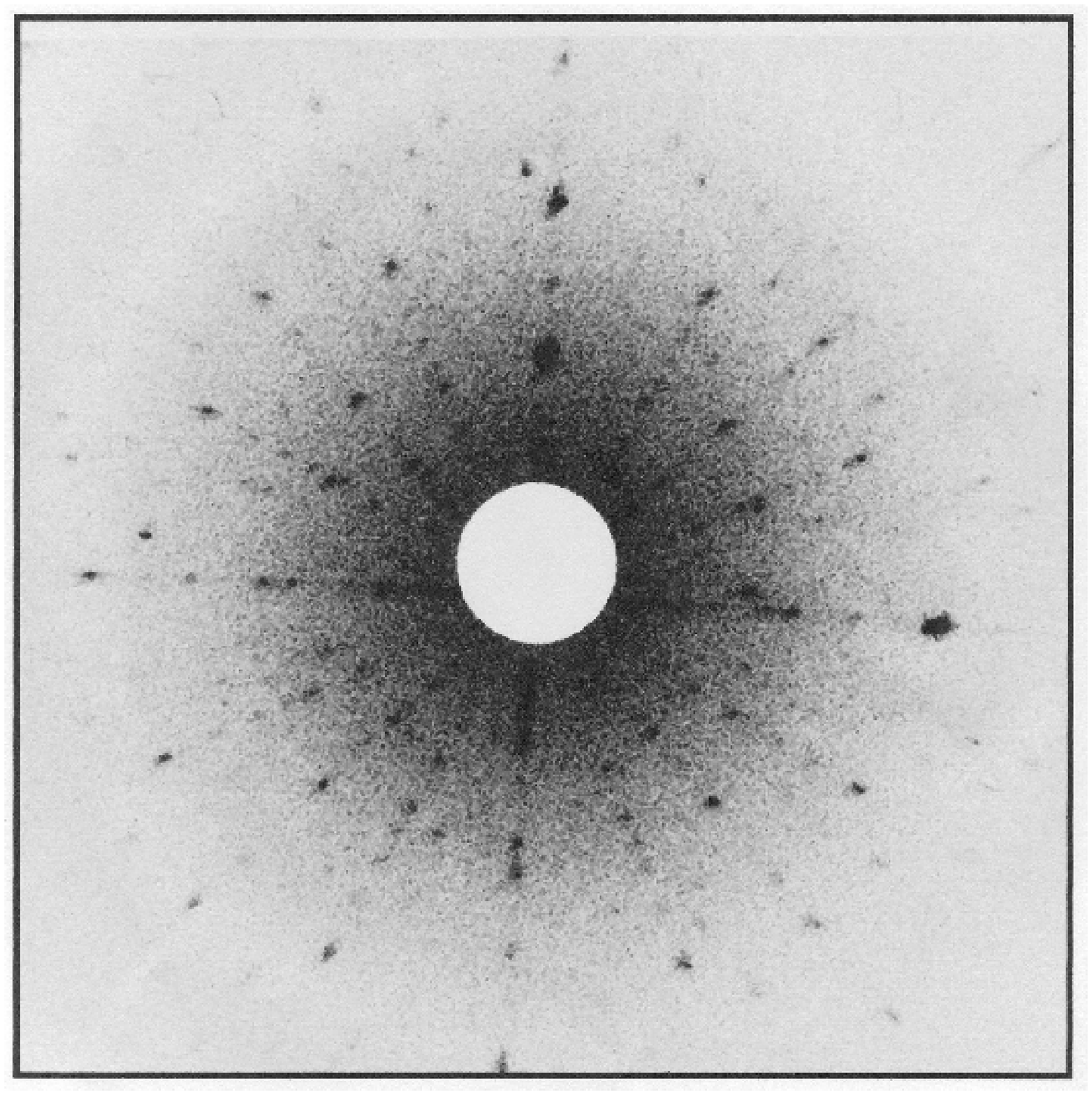}
\caption{\label{fig1}Laue photograph of single crystal of composition
 x=0.67.}
\end{figure}

To show the quality of the single crystal we have taken Laue x-ray diffraction pictures on the cleaved surface of the NCMO
single crystals. Fig.~\ref{fig1} is given as an illustrative example,
it can be clearly observed from this figure that there are clear spots
and the expected four-fold symmetry in the (001) plane.
We recall that the Space Group [SG] of NCMO-214 system is
Bmab [SG number 64] above 220 K [roughly] and is Pccn [SG number 52]
below 220 K, for both x=0.67 and x=0.70. For the composition x=0.60
below 160 K (roughly) SG of the fundamental structure: Pccn (56)
and above 160K (roughly) SG of the fundamental structure: Bmab (64)

     The HC of the sample is calculated by subtracting
the addenda measurement from the total heat capcity measurement. The
total HC is the measurement of the HC of the sample, the grease,
and the sample platform. The two measurements-one with and one without
the sample on the sample platform are necessary for accuracy. In order
to ensure the further accuracy of our results, we conducted the
experiment several times, each time repeating the addenda measurement.
In addition, our several independent measurements are separated
by 20-30 days.We note that automatic subtraction of the addenda, at each
sample temperature measurement is performed. As we were interested mainly
in the high temperature region, we used H-grease.

The dc magnetization measurements were performed with a SQUID
magnetometer (MPMSXL, Quantum Design). Zero-field cooling (ZFC)
measurements were carried out in the following way: first at zero field
cooling down temperature from 390 to 5 K, then applying a field of 10000 Oe,
and finally measuring the sample from 5 to 390 K in the field.
Field cooling (FC) measurements were performed immediately after ZFC measurement from 390 to 5 K under 10000 Oe.


\section{Results and Discussion}
\begin{figure}
\includegraphics[width=6in,height=6in]{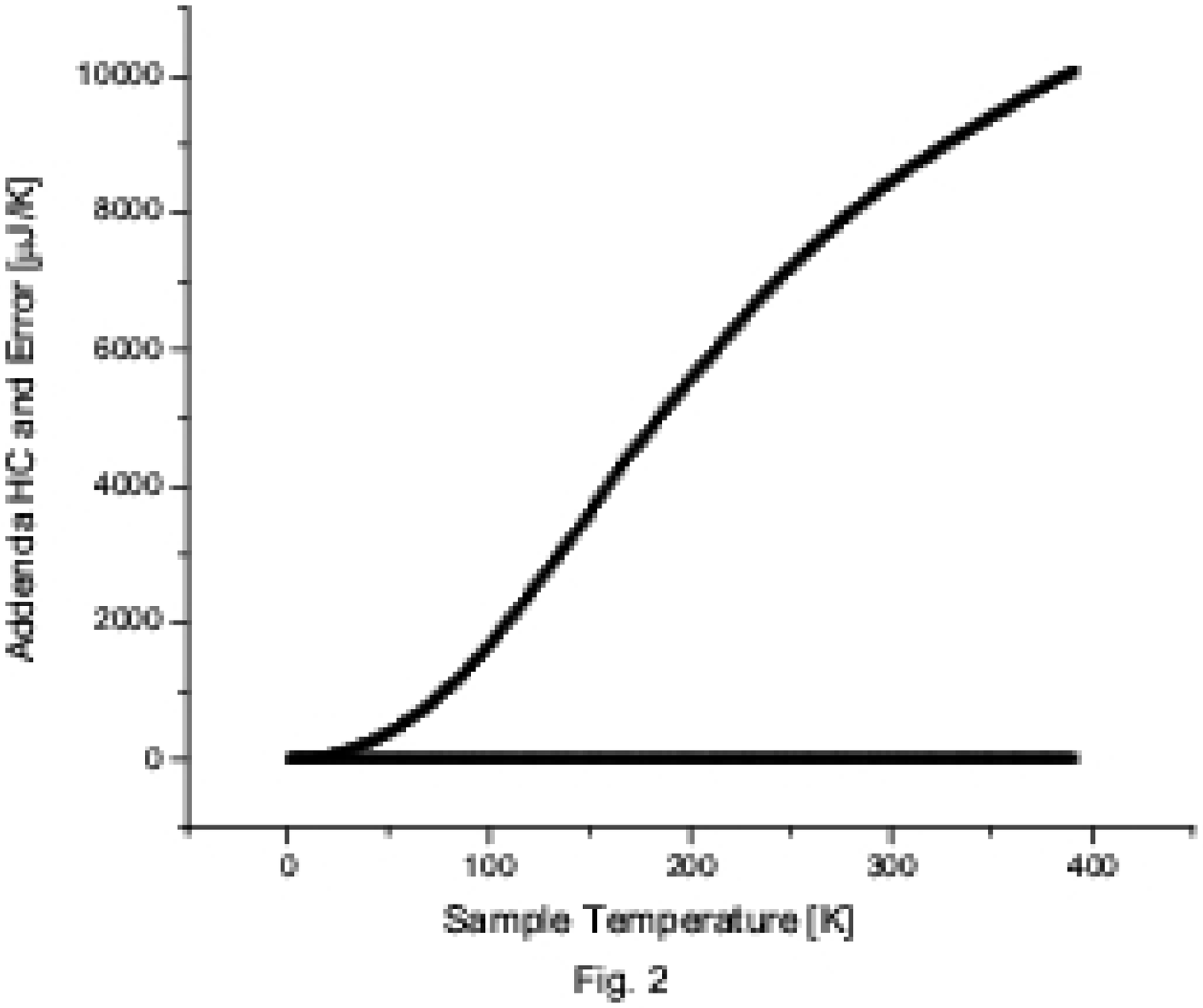}
\caption{The Addenda HC [$\mu$ J/K] used for the $x=0.67$
sample.}
\label{fig2}
\end{figure}

\begin{figure}
\includegraphics[width=6in,height=6in]{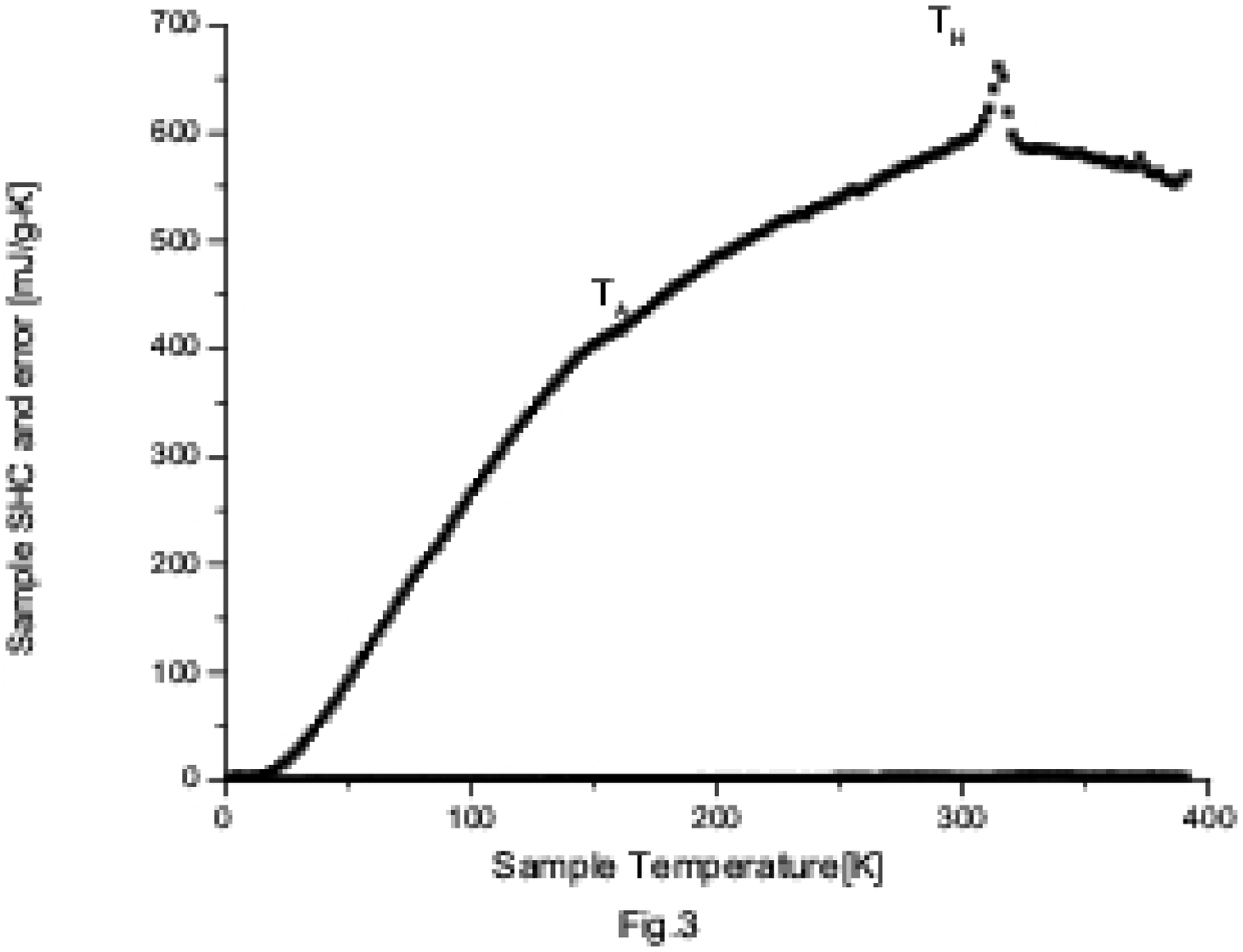}
\caption{The sample HC and the corresponding error for the case
$x=0.67$.}
\label{fig3}
\end{figure}

\begin{figure}
\includegraphics[width=6in,height=6in]{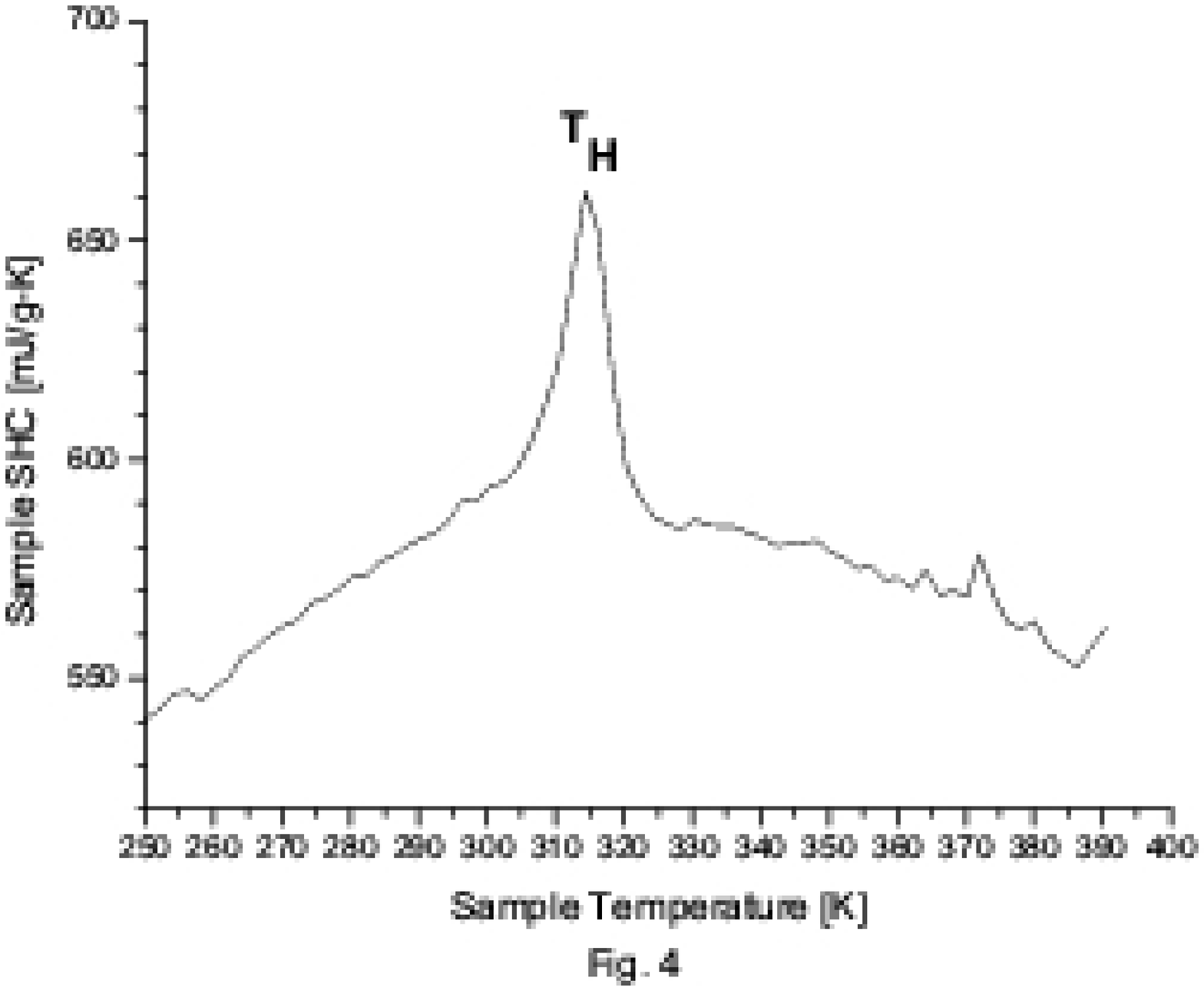}
\caption{The main HC anomaly in more detail
for the case of $x=0.67$.}
\label{fig4}
\end{figure}

\begin{figure}
\includegraphics[width=6in,height=6in]{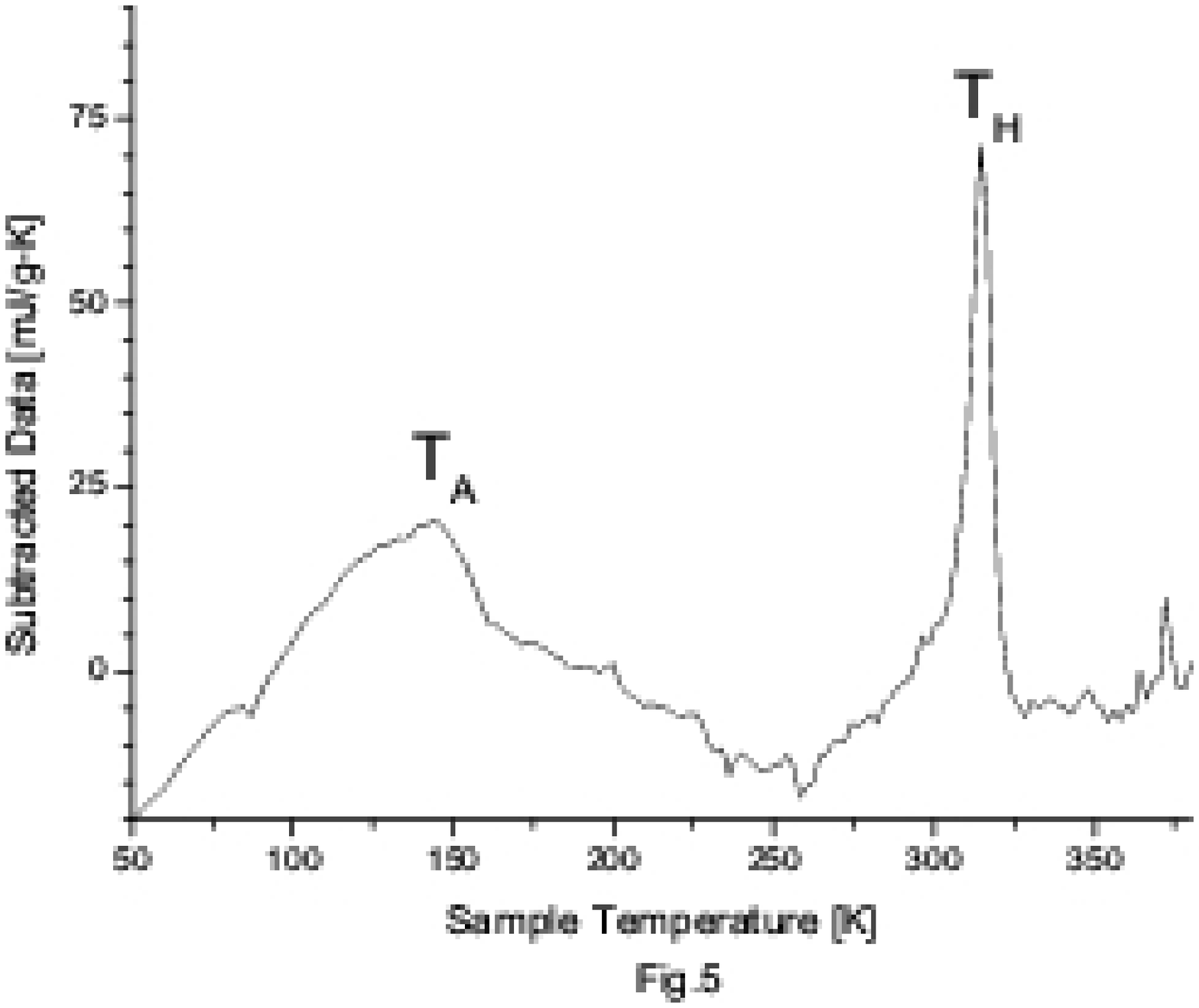}
\caption{The subtracted data, showing the anomalies at
T$_{_H}$  and T$_{_A}$ for the sample with $x=0.67$.}
\label{fig5}
\end{figure}

\begin{figure}
\includegraphics[width=6in,height=6in]{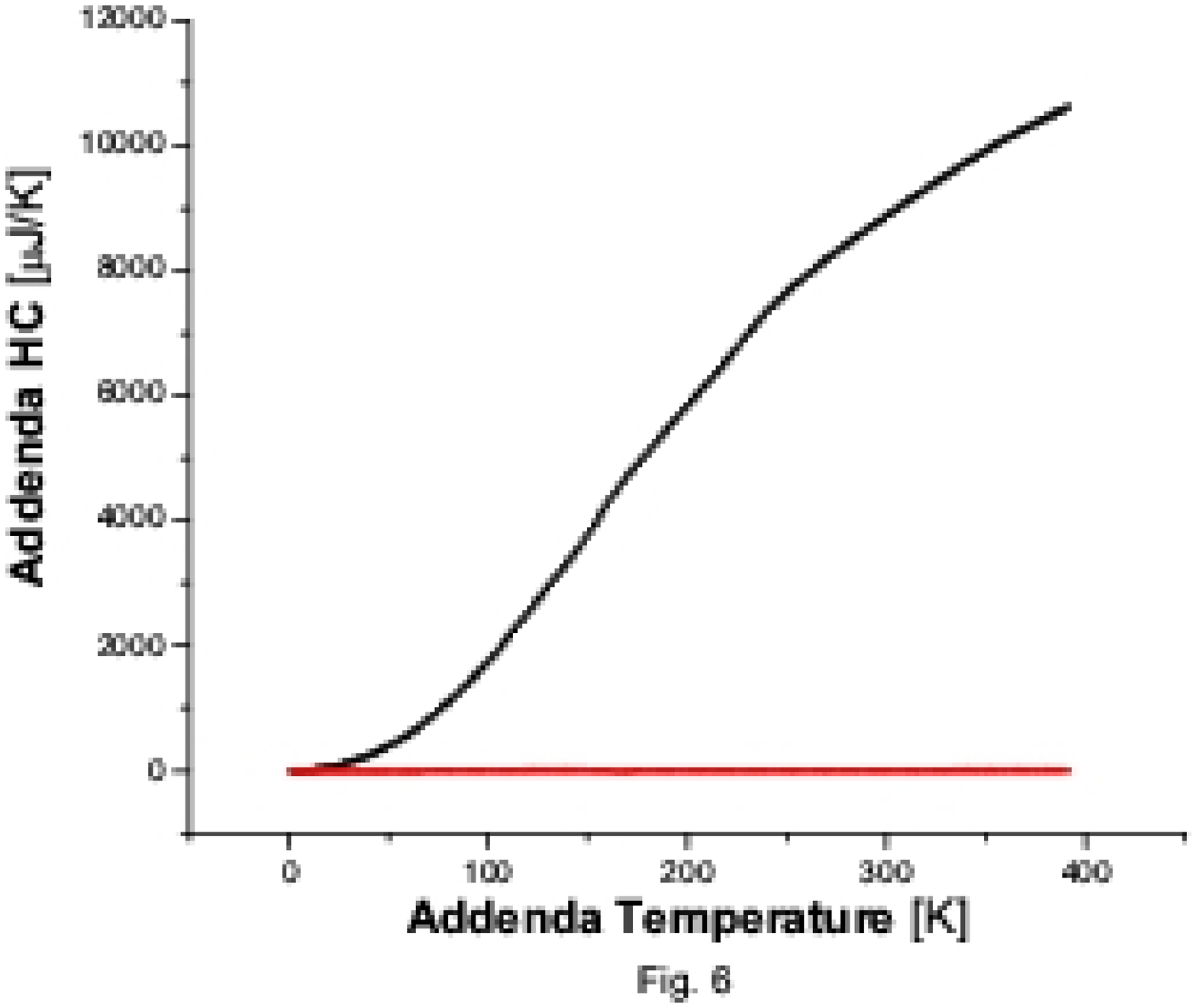}
\caption{Results for the HC [$\mu$ J/K] of the Addenda and
the error which was used for the $x=0.70$ sample.}
\label{fig6}
\end{figure}

\begin{figure}
\includegraphics[width=6in,height=6in]{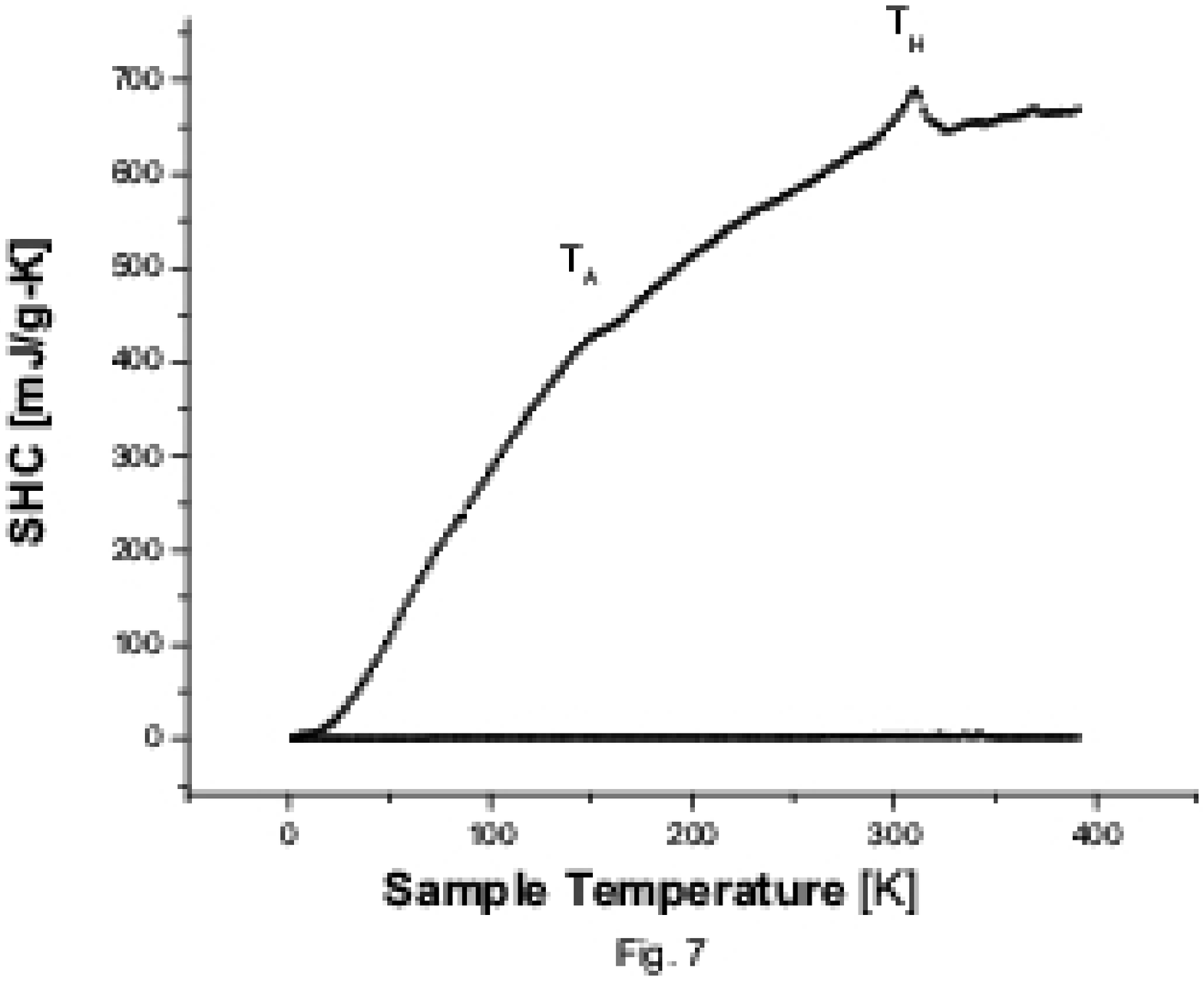}
\caption{The sample HC and the corresponding error for the case
when $x=0.70$.}
\label{fig7}
\end{figure}

\begin{figure}
\includegraphics[width=6in,height=6in]{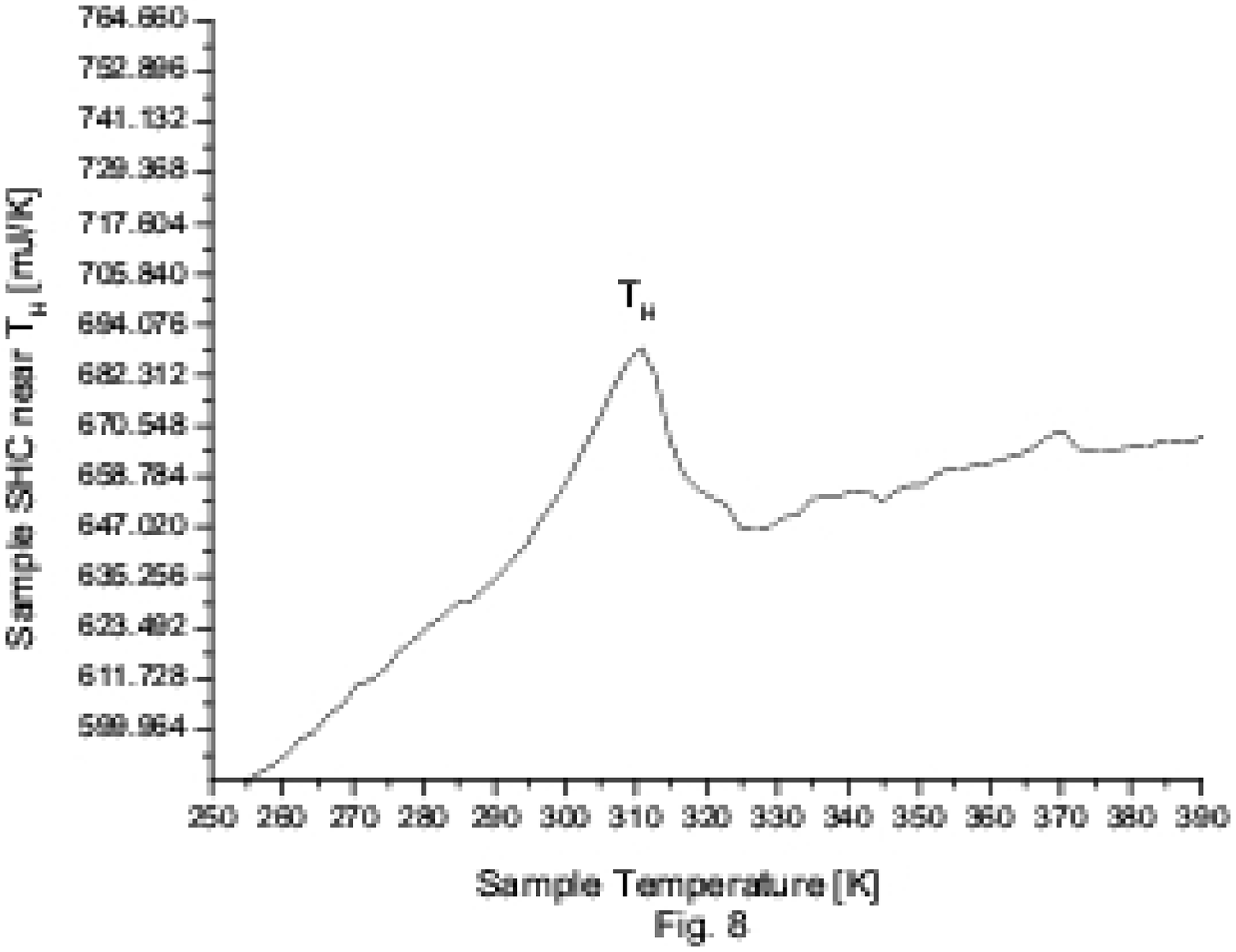}
\caption{The main HC anomaly in more detail
for the case of $x=0.70$.}
\label{fig8}
\end{figure}

\begin{figure}
\includegraphics[width=6in,height=6in]{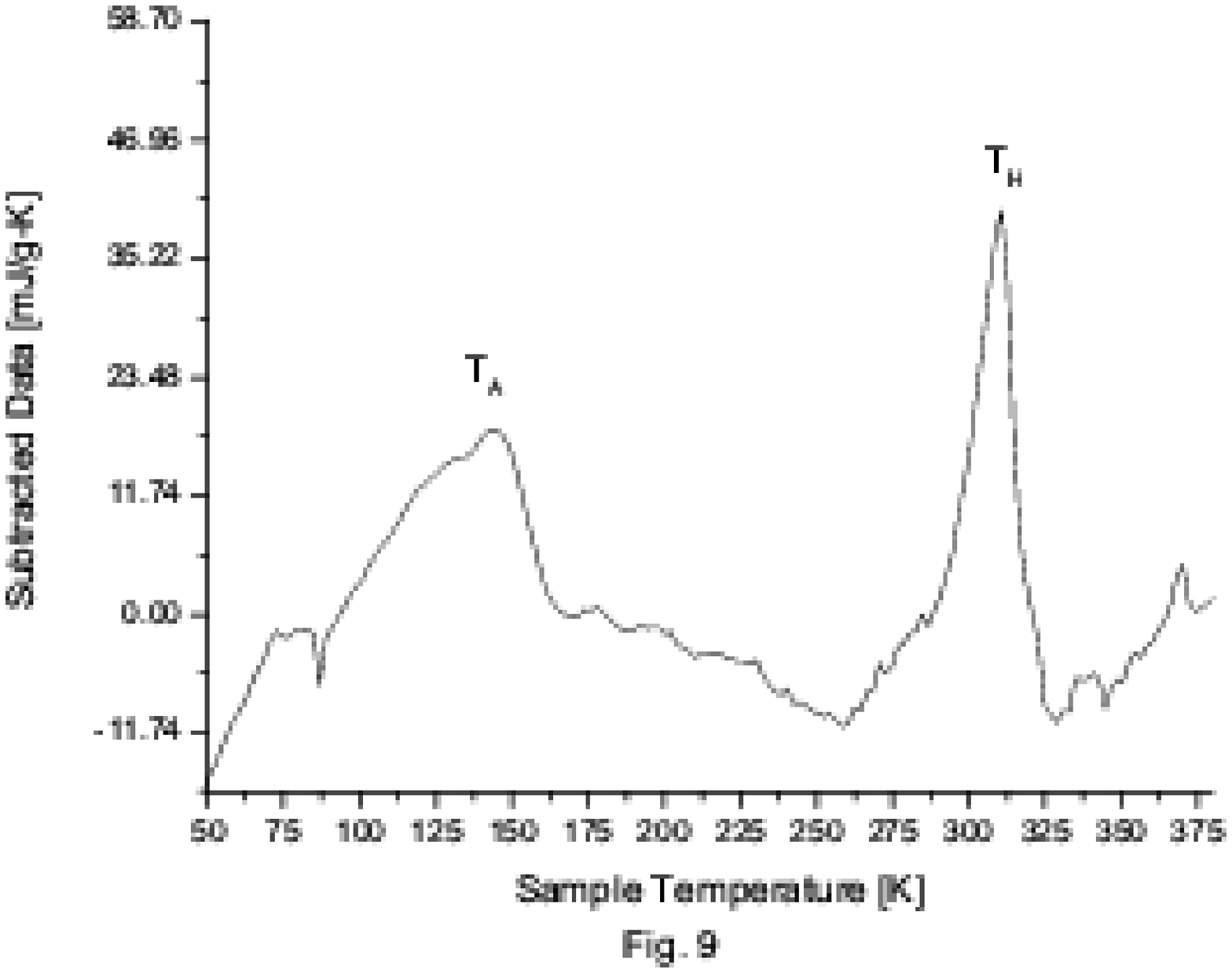}
\caption{The subtracted data, showing the anomalies at
T$_{_H}$  and T$_{_A}$ for the sample with $x=0.70$.}
\label{fig9}
\end{figure}

Figs.~\ref{fig2}-\ref{fig5} show the results of HC measurements
for the x=0.67 sample. Similarly the results are displayed in
Figs.~\ref{fig6}-\ref{fig9} when the composition is x=0.70. Let
us call these cases I and II respectively. We
note that both data are self-consistent, after one takes into
account the simple factor due to the mass of the sample. The
mass of the sample in case I was measured to be 5.25 $\pm$ 0.10 mg,
and in case II 8.34 $\pm$ 0.10 mg. The clarity of the measurement
is vividly displayed in Figs.~\ref{fig4} and \ref{fig8}. We can
identify two main deviations from the normal at temperatures
 $T_{_H}$ and  $T_{_A}$ in Figs.~\ref{fig3} and \ref{fig7}.
The magnitude of the peaks is 5-15 \% above the 'normal' background
in the corresponding temperature range.

       To see the anomalous peaks more clearly, we subtract
the smooth background, and the results are displayed in Figs.~\ref{fig5}
and \ref{fig9}, respectively for the cases I and II. We label
this deviation in Heat Capacity  as $\Delta H$. The location
of the peaks is at $T_{_A}^{HC}=143.29$ K,  $T_{_H}^{HC}=314.35$ K for the case I, and in case II we find $T_{_A}^{HC}=143.47$ K,  $T_{_H}^{HC}=310.73$ K. From the peaks at $T_{_H}^{HC}$ we find a percentage change, or deviation away from the background value, of 12.931 \% and 8.125 \% for cases I and II respectively. The peaks at $T_{_A}^{HC}$ represent a change of approximately 11.764 \% for case I, and a 8.043 \% departure from the background value, for case II.

       The origin of the strong and distinct anomalous peak in the
Heat Capacity at T$_{_H}$ is due to charge-orbital ordering, which supports
our previous work on charge and orbital ordering using electron
microscopy \cite{nag03}. One can also regard this as more direct
evidence for transverse and sinusoidal structural modulations.
It is also tempting to suggest that this feature is related to
a charge-orbital wave of e$_g$ electrons in this material.
The alteration in the manganese valence is nothing but the
variation of the density of e$_g$ electrons. The possible reason
for the modulation of the manganese valence, is the successive change
in the amplitude of the Jahn-Teller distortion in the MnO$_6$ octahedra
with position \cite{nag02, lar01}. It is also possible to interpret
the modulated structure as an orbital density wave, as mentioned
previously\cite {nag03}. The density of e$_g$ electrons is constant
in a pure orbital density wave, however in our case we assume that
orbital state varies as shown in Fig.~6a of \cite{nag03} between $\pm \pi$.
It is useful to recall the description of orbital state by
the pseudo-spin space\cite{kan60, kug73}. It is assumed that the
motion of the pseudo-spin is confined to the xz plane and orbital state at the
site i, $\theta_i$varies according to,
\begin{eqnarray}
\theta_i &=&\cos(\theta_i/2)|x^2-y^2> +\sin(\theta_i/2)|3z^2-r^2>.\nonumber
\end{eqnarray}
If it is assumed that a pure charge density wave, then the variation
of the orbital state in the direction perpendicular to the stripe
is discrete taking on values  $\pm \pi$. In contrast, the Mn valence
or $e_g$ electron density varies in a sinusoidal manner. Thus it
is more tempting to assume the variation of orbital-density wave
to follow the variation of the charge density wave, as also
proposed in \cite{koi98}. In brief we take the variation of both
charge density and orbital density waves to be sinusoidal.
Incidentally, we can use this interpretation of the sinuosoidal
variation of charge-orbital density to interpret the anomaly at T$_{_A}$.
We may simply see that in this temperature region a further
 alteration occurs in the charge and magnetic state of the material,
perhaps involving a further change in the local structure, arising
from a weak Jahn-Teller effect.

\begin{figure}
\includegraphics[width=6in,height=6in]{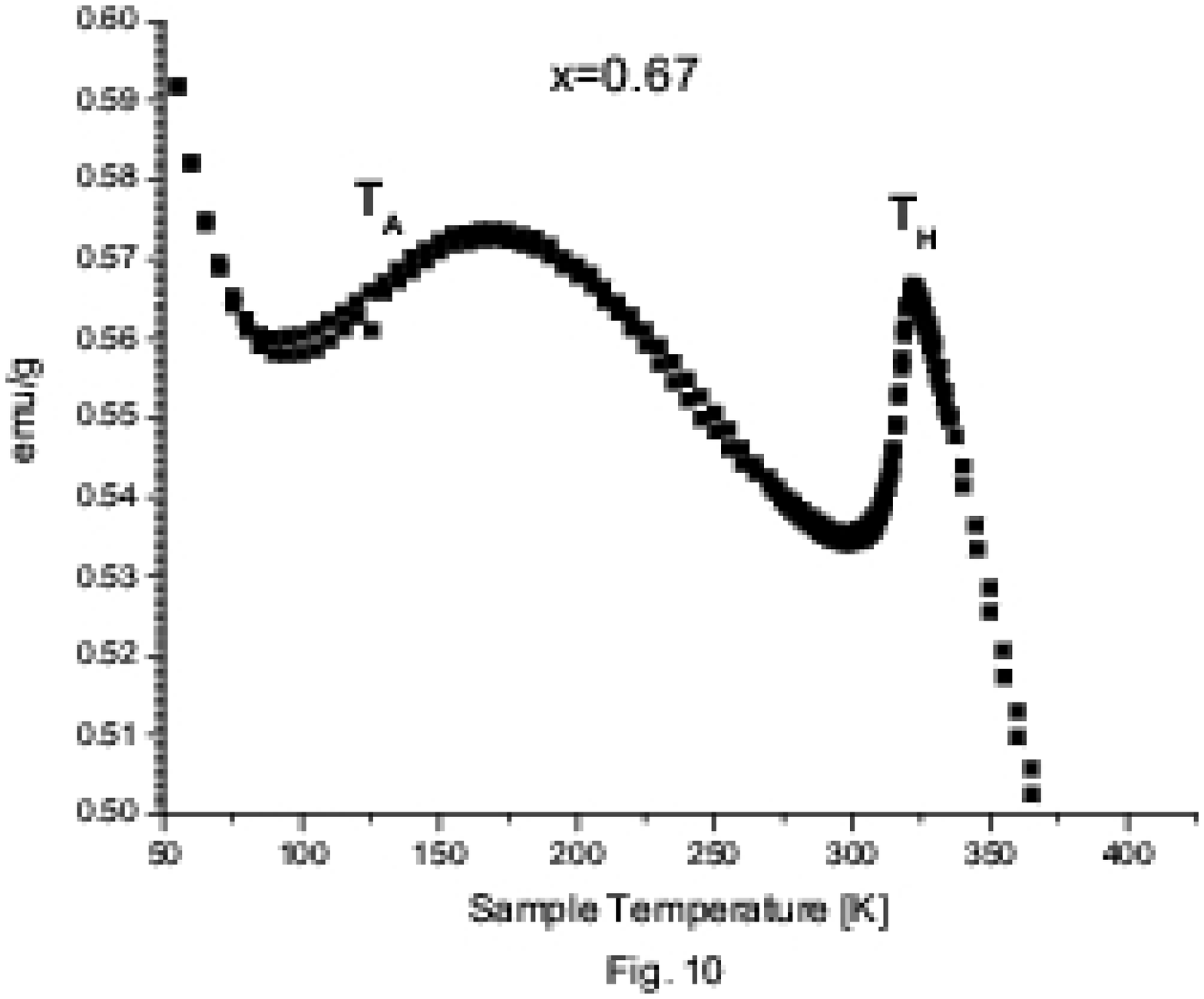}
\caption{\label{fig10}The Long Moment [emu/g] for the composition
 x=0.67, showing clearly the anomalies at T$_{_H}$ and T$_{_A}$.The applied field is 1T. }
\end{figure}

\begin{figure}
\includegraphics[width=6in,height=6in]{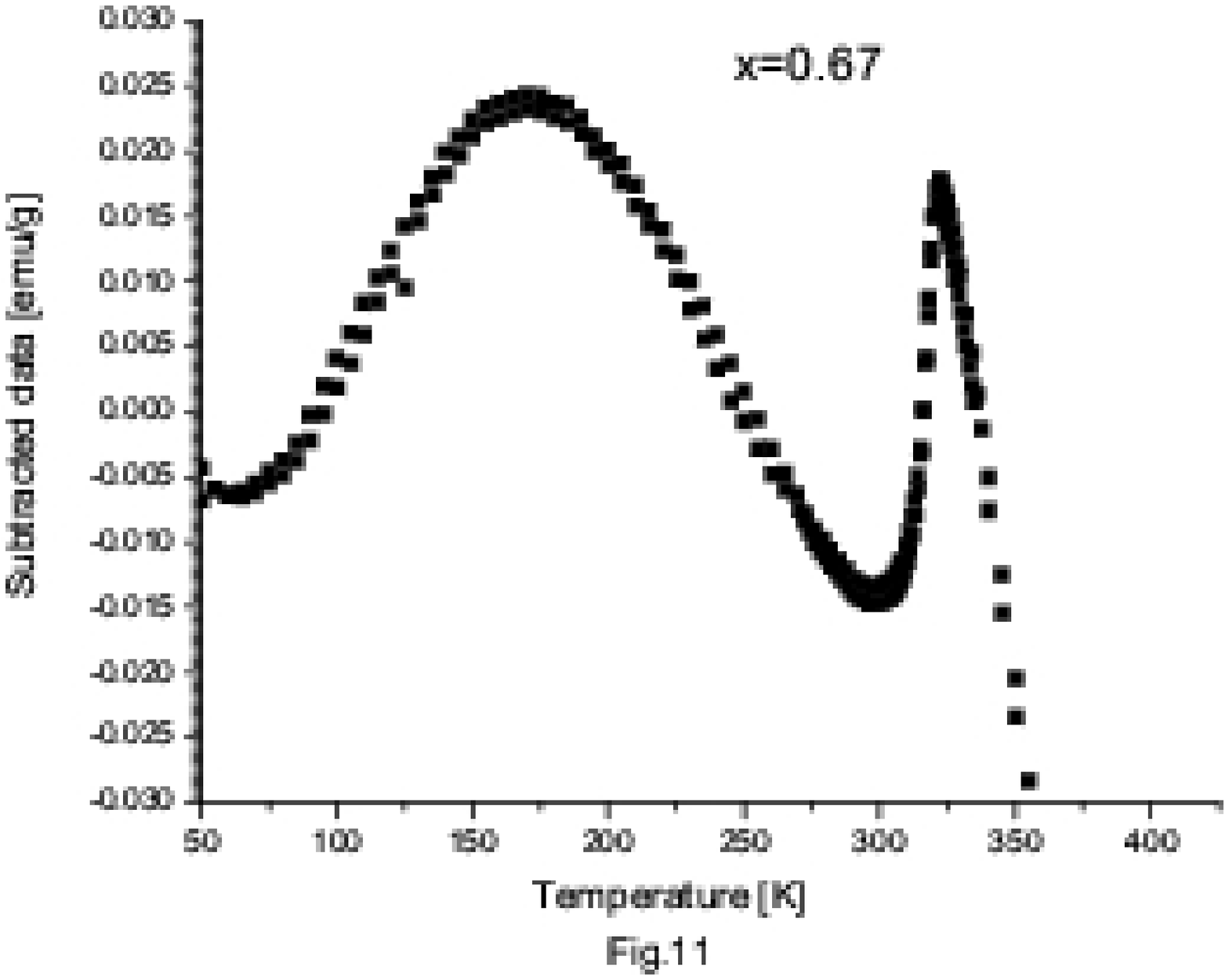}
\caption{\label{fig11}The subtracted data [emu/g] for the composition
 x=0.67, showing clearly the anomalies at T$_{_H}$ and T$_{_A}$.The applied field is 1T. }
\end{figure}

    The results of the dc magnetic measurements are displayed in
Figs.~\ref{fig10}-\ref{fig11} for cases I and II respectively.
To see the anomalous peaks more clearly, we subtract
the smooth background, and the results are displayed in Figs.~\ref{fig11}
and \ref{fig13}, respectively for the cases I and II. We label
this deviation in Heat Capacity  as $\Delta M$. The location
of the peaks is at $T_{_A}^{M}=164.76$ K,  and $T_{_H}^{M}=323.00$ K for the case I,
and in case II we find $T_{_A}^{M}=164.86$ K,  $T_{_H}^{M}=318.73$ K. From the
peaks at $T_{_H}^{M}$ we find a percentage change, or deviation away from
the background value, of 7.69 \% and 6.45 \% for cases I and II respectively.
The peaks at $T_{_A}^{M}$ represent a change of approximately 5.59 \% for case I,
and a 3.30 \% departure from the background value, for case II.

It may be remarked that as our simulations of electron
microscopy \cite{nag03} are consistent with
the Wigner-Crystal model, it is tempting to interpret our HC data
in this context by thinking of it as signalling the formation
of Wigner-Crystal at temperature T$_{_H}$. At the temperature
T$_{_A}$, the anomaly may be seen to arise from the charge-orbital
ordering alteration of the magnetic state of the material. The
only work we located for which similar interpretation was given
is \cite{mah03} for the case of cobaltate polycrystalline
Pr$_{0.5}$Co$_{0.5}$ O$_{3}$. It is interesting to note, that two
anomalies have been reported for the case of the cobaltate
polycrystalline Pr$_{0.5}$Co$_{0.5}$ O$_{3}$,
\cite{mah03}, see Fig. 4 of \cite{mah03}. The anomaly at T$_{_A}$ is
attributed to the orbital-ordering alteration of the ferromagnetic state.

    In order to further understand the difference between commensurate and
incommensurate behavior, we present the ED diffraction pattern for
case x=0.67 and x=0.60. Fig.~\ref{fig14} and ~\ref{fig15} show ED
patterns for x=0.67 at 80 K and 290 K respectively.
Likewise, Fig.~\ref{fig16} and ~\ref{fig17} show ED patterns for
the case x=0.6 at 80 K and 290 K.
In Fig.~\ref{fig16}, super-lattice reflections due to charge-orbital ordering are indicated by black arrows. The hk0 reflections (h,k:odd)
indicated by white arrow are the fundamental reflections of low-temperature
Pccn(56) structure. The reflections are invisible in Fig.~\ref{fig15}, indicating the structural phase transition to high-temperature Bmab(64) structure, as already mentioned. However, the super-lattice reflections due to charge-orbital ordering are still clearly visible.

In Fig.~\ref{fig16}, the hk0 reflections are very weak. In addition,
super-lattice reflections are invisible at 290K as seen in Fig.~\ref{fig17}.
From the ED experiments, it is roughly estimated that the temperature
at which the super-lattice reflections appear is maximum around x=0.67.
The temperature at which the hk0 reflections appear also might be maximum around x=0.67.

These observations are supported by the HC and dc magnetization studies, as
indicated above. The HC and magnetization data for x=0.60 indicates that
transitions are weaker as compared to x=0.67.

Low-temperature electron diffraction measurements confirmed super-lattice reflections due to the charge-orbital ordered (COO) states at doping levels of $0.55\leq x \leq 0.75$ for Nd$_{1-x}$Ca$_{1+x}$MnO$_4$ system. High-resolution electron microscopy observations revealed transverse and sinusoidal structural distortions with a doping-level-dependent period of $d_s=a/(1-x)$ in the COO phases. The d$_s$ is commensurate to the lattice period at commensurate doping levels, e.g. 3a at x=0.67, or 4a at x=0.75, and incommensurate at incommensurate doping levels, e.g. 3.3a at x=0.7. The successive Jahn-Teller distortions imply successive modulations of the e$_g$ orbital state and the Mn valence with the period of d$_s$. From the EM we can see, in terms of the COO states, a small but significant difference between commensurate and incommensurate doping levels. If we take the heat capacity and the dc magnetization results in addition to the EM results we can clearly see a difference between the commensurate and incommensurate dopings.
\section{Conclusions}
We have for the first time provided clear experimental
evidence for charge-orbital ordering and related transition using
Heat Capacity and dc magnetization measurements in the material
Nd$_{1-x}$Ca$_{1+x}$MnO$_{4}$. This supports the previous HRTEM
and ED experimental work of two of the authors \cite{nag03}, which gave
evidence of the presence of a charge-orbital density wave
in layered manganites Nd$_{1-x}$Ca$_{1+x}$MnO$_{4}$.
The measurements suggest that the region near and of the anomaly
is  non-perturbative. In addition, we report for the first time the
growth of single crystals of  Nd$_{1-x}$Ca$_{1+x}$MnO$_{4}$, to our
knowledge this has not been reported before.
\section*{Acknowledgments}
The Sher Alam's work is supported by the Japan Society for
for Technology [JST] through the STA fellowship and MONBUSHO
via the JSPS invitation program.

\newpage
\begin{figure}
\includegraphics[width=6in,height=6in]{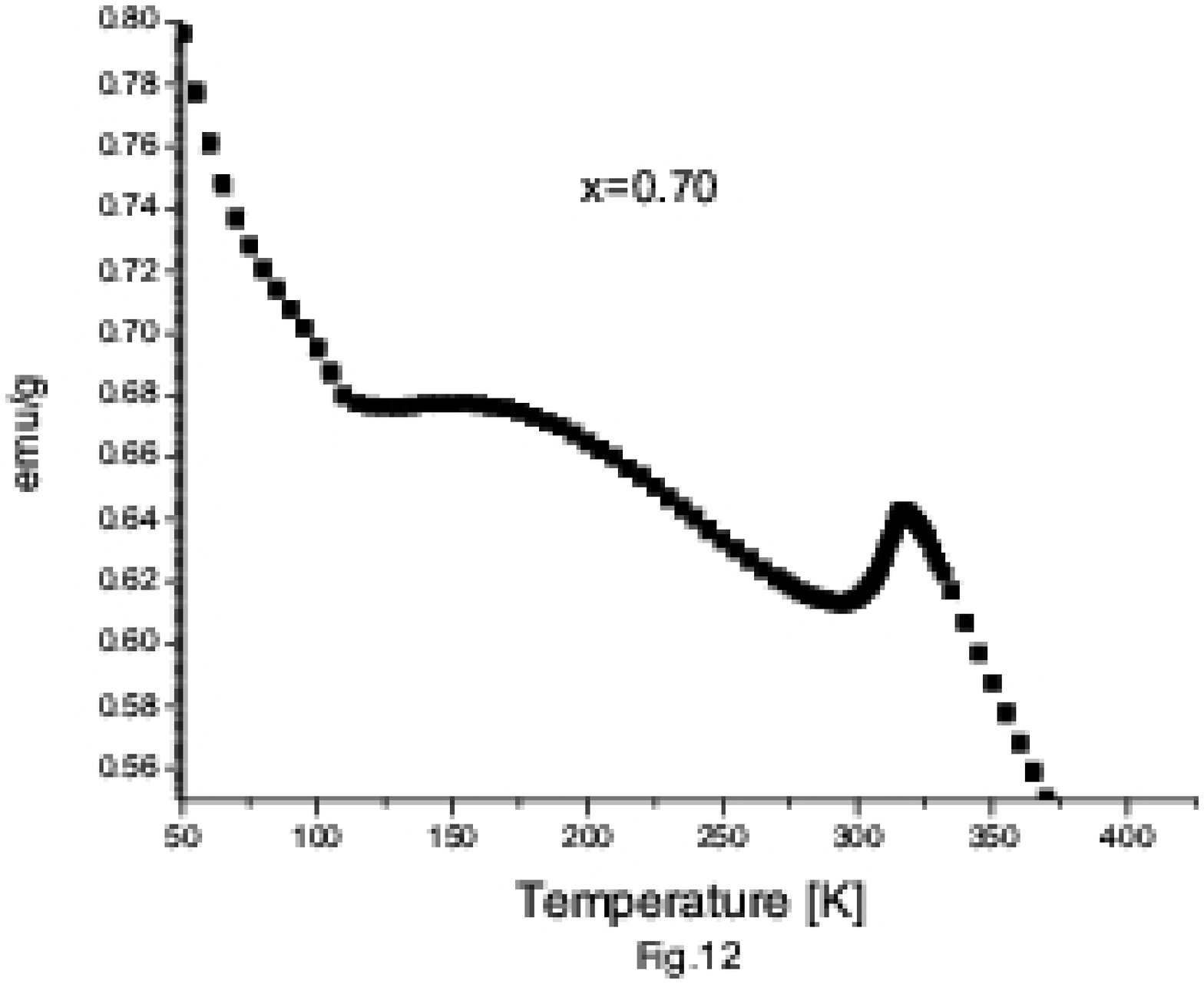}
\caption{\label{fig12}The Long Moment [emu/g] for the composition
 x=0.70, showing clearly the anomalies at T$_{_H}$ and T$_{_A}$. The applied field is 1T.}
\end{figure}
\begin{figure}
\includegraphics[width=6in,height=6in]{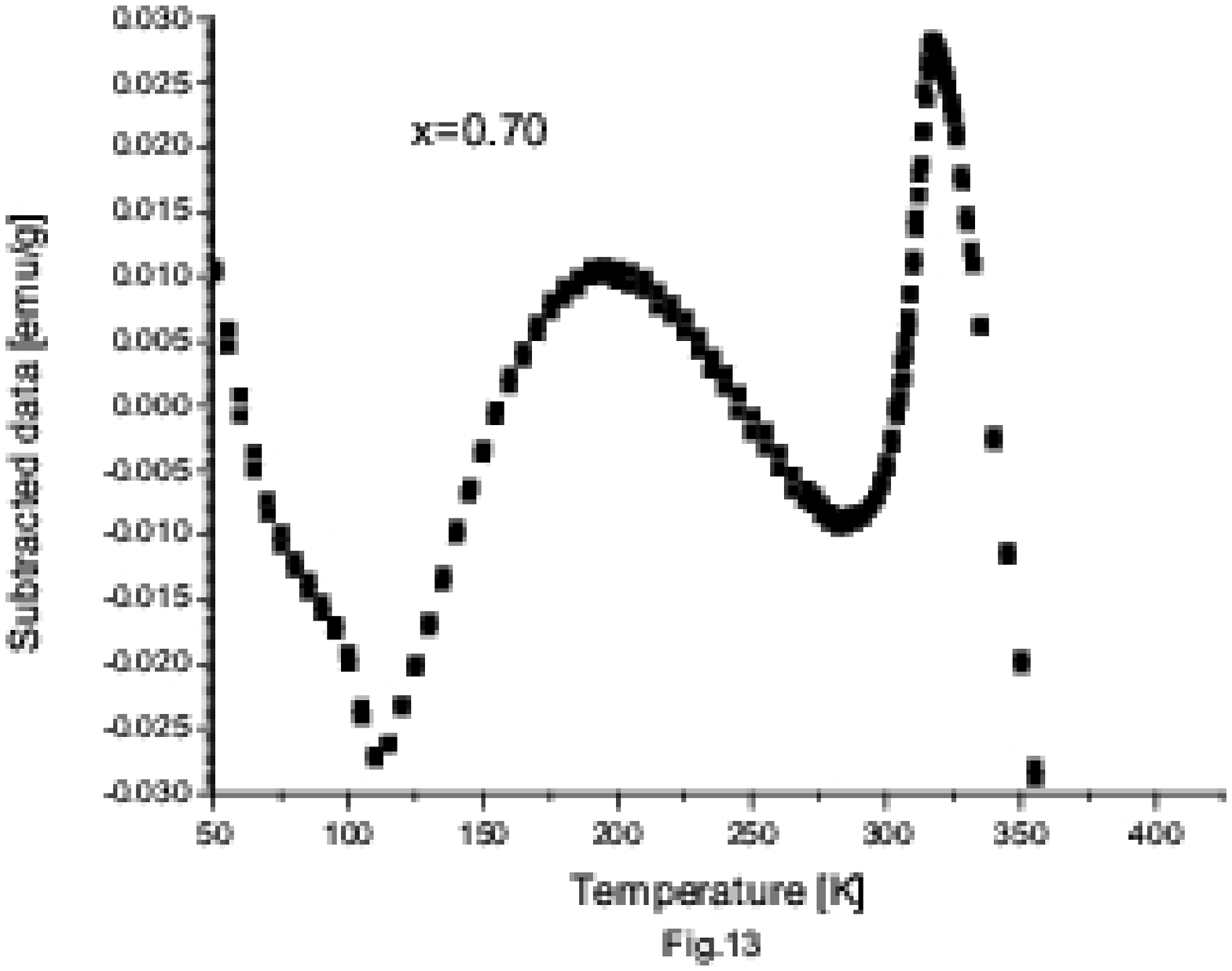}
\caption{\label{fig13}The subtracted data [emu/g] for the composition
 x=0.70, showing clearly the anomalies at T$_{_H}$ and T$_{_A}$.The applied field is 1T. }
\end{figure}
\begin{figure}
\includegraphics[width=6in,height=6in]{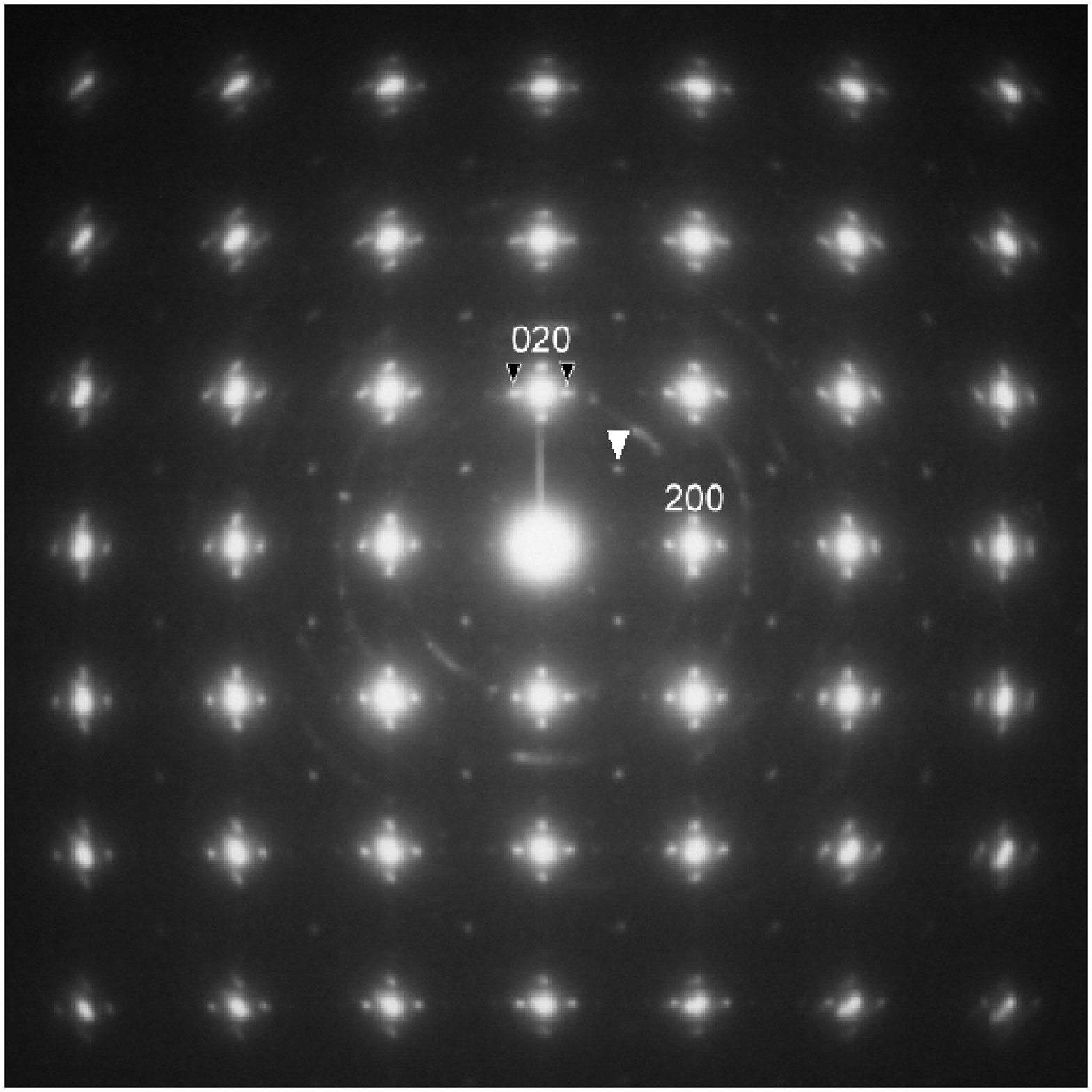}
\caption{\label{fig14}ED pattern for the composition x=0.67 at 80 K.}
\end{figure}
\begin{figure}
\includegraphics[width=6in,height=6in]{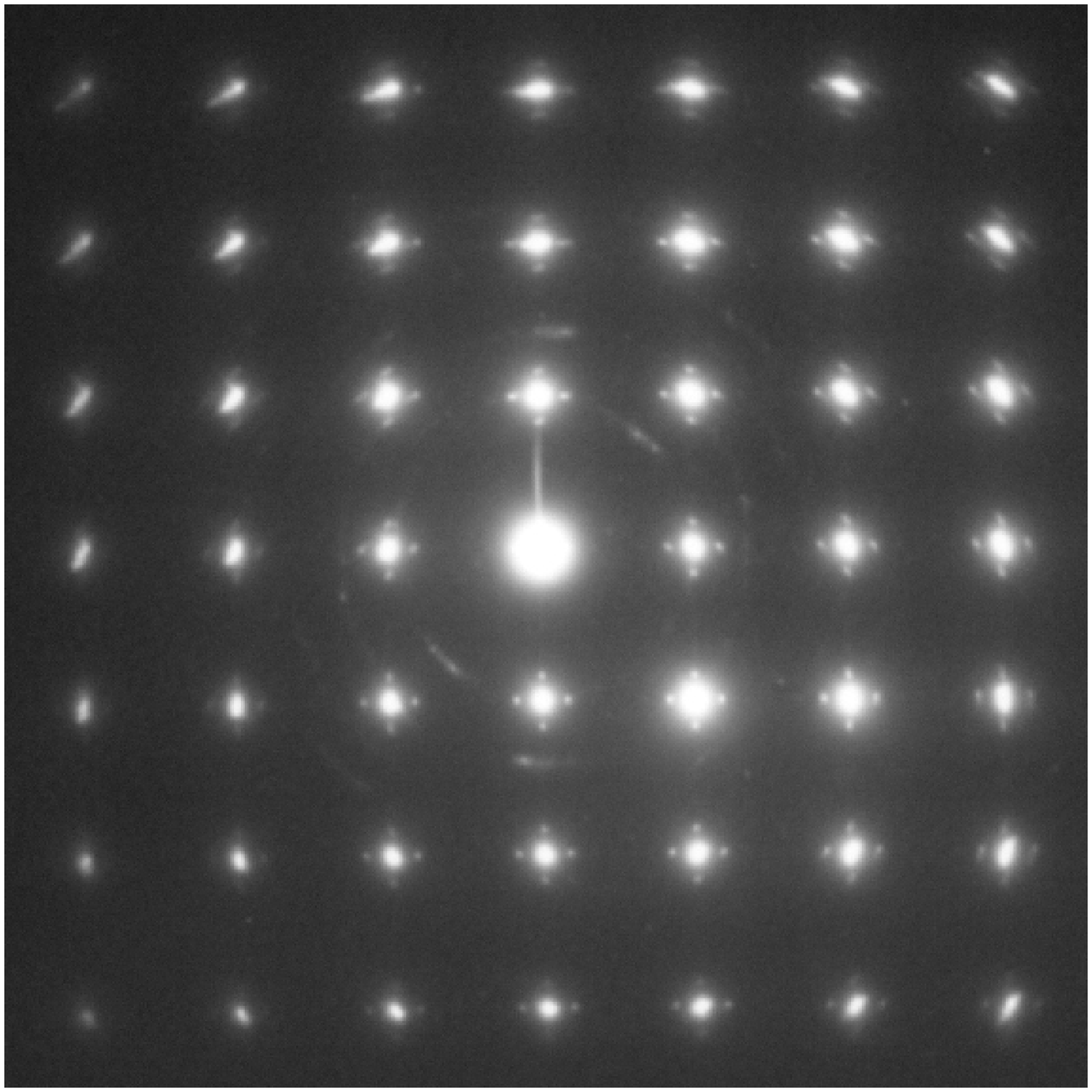}
\caption{\label{fig15}ED pattern for the composition x=0.67 at 290 K.}
\end{figure}
\begin{figure}
\includegraphics[width=6in,height=6in]{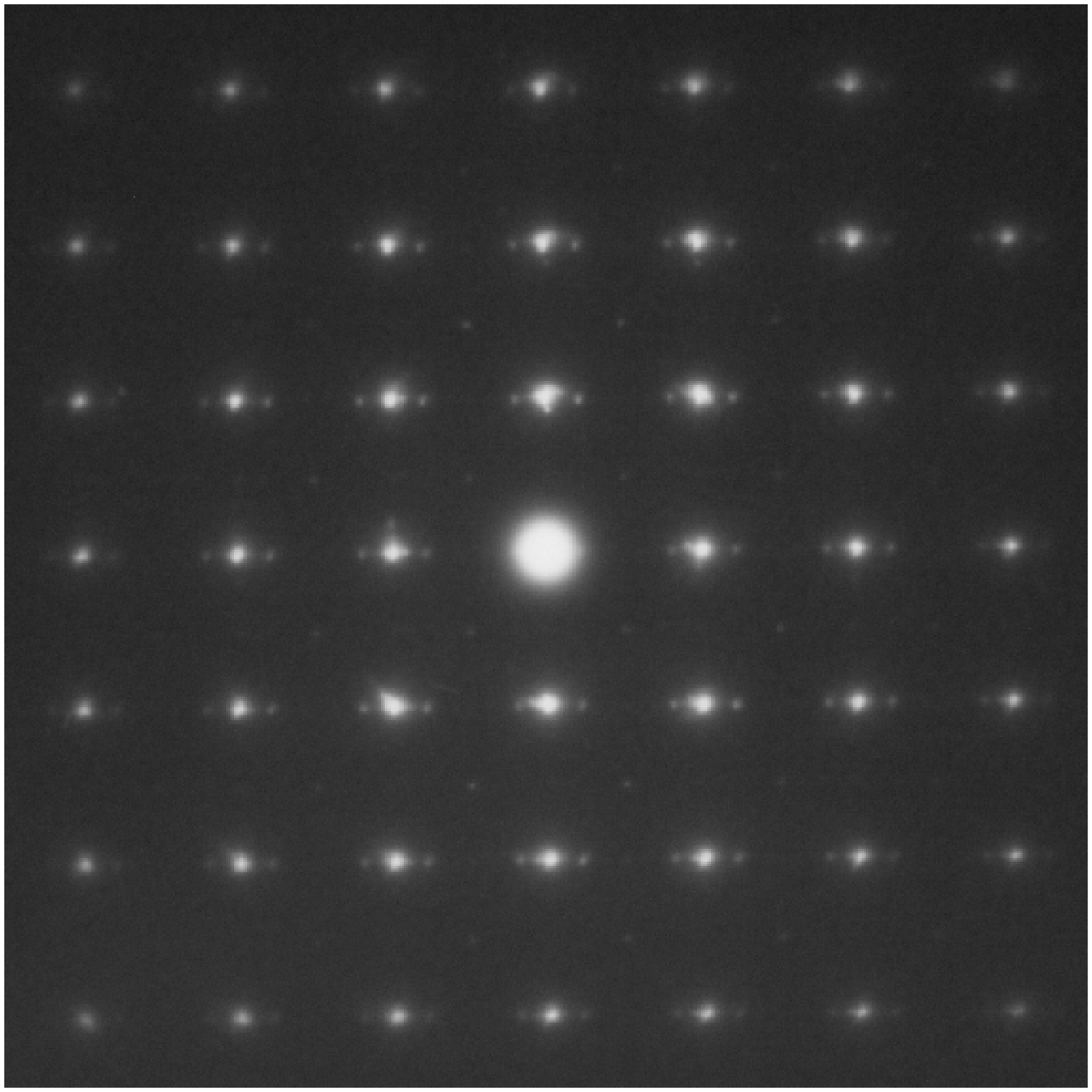}
\caption{\label{fig16}ED pattern for the composition x=0.60 at 80 K.}
\end{figure}
\begin{figure}
\includegraphics[width=6in,height=6in]{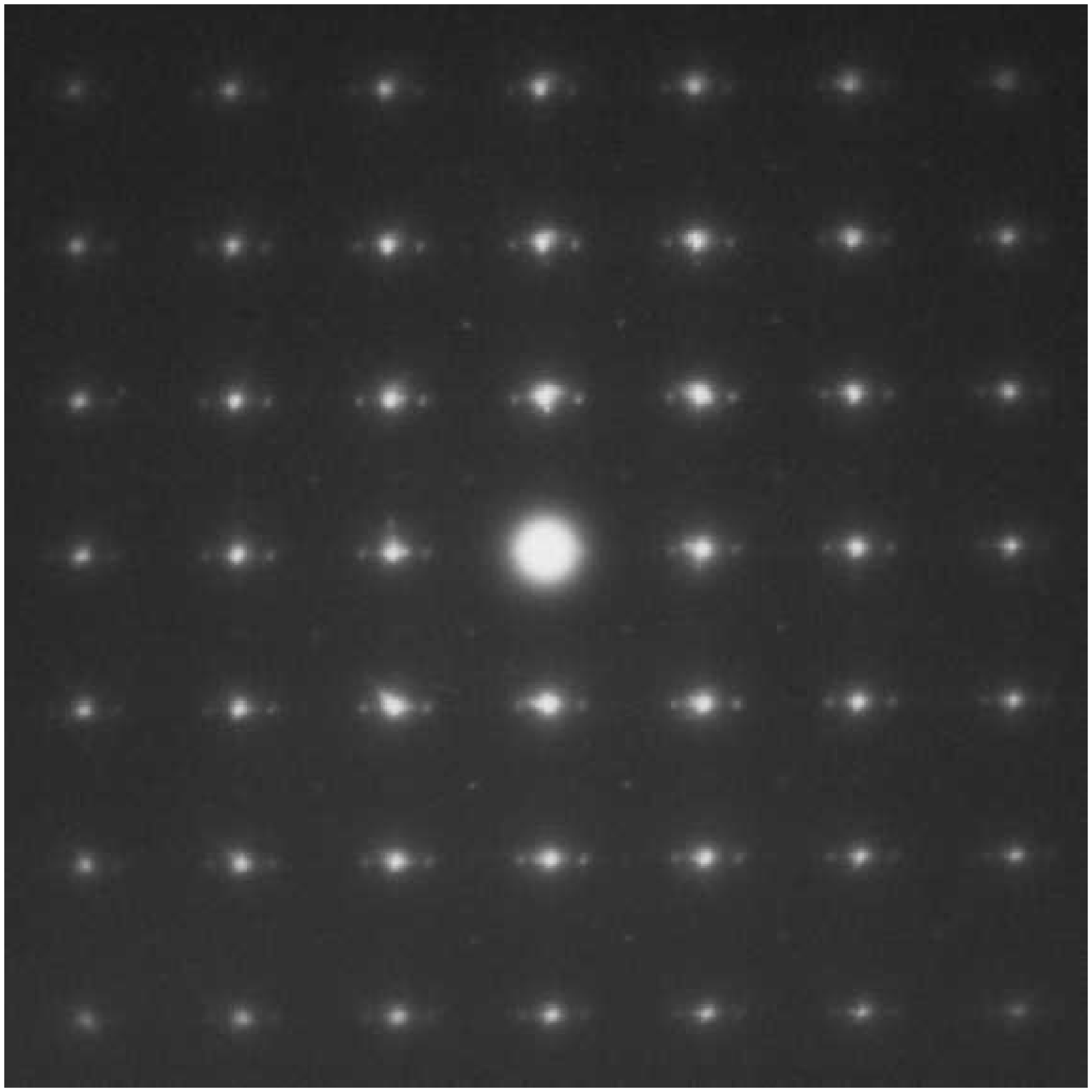}
\caption{\label{fig17}ED pattern for the composition x=0.60 at 290 K.}
\end{figure}
\end{document}